\documentclass[12pt]{article} 
\usepackage[latin1]{inputenc}
\usepackage [dvips]{graphicx}
\usepackage{cite} 
\newcommand{\ee}{\end{equation}}
\newcommand{\be}{\begin{equation}}
\begin{document}

\title {Partial energies fluctuations and negative heat capacities}
\author{X. Campi\thanks{Corresponding author: campi@ipno.in2p3.fr}  and H.
Krivine \\ \small Laboratoire de Physique Th\'eorique et Mod\`eles
Statistiques \thanks{Unit\'e de Recherche de l'Universit\'e de Paris XI
associ\'ee au CNRS (UMR 8626)}\\ \small B\^at. 100, Universit\'e de Paris
XI, F-91405 Orsay Cedex, France \\ \\ E. Plagnol \\ \small APC-PCC/Collège de
France, F-75231 Paris Cedex 05, France \\ \\ N. Sator \\ \small LPTL,
Université de Paris VI, 4 Place Jussieu F-75252 Paris Cedex 05, France}

\maketitle

\it Abstract. We proceed to a critical examination of the method used in
nuclear fragmentation to exhibit signals of negative heat capacity. We show
that this method leads to unsatisfactory results when applied to a simple
and well controlled model. Discrepancies are due to incomplete evaluation of
potential energies.  \rm

\vskip 1cm 

The possibility to observe negative heat capacities has been recently the
object of much interest and debate
\cite{GROSS,LABASTIE,CHOMAZ,GULMINELLI,PLEIMLING-HUELLER,DORSO,DAGOSTINO,DAGOSTINO2,SCHMIDT,FARIZON}. This
is partly because a negative heat capacity seems counter-intuitive (how can
adding energy to a system make it cooler ?) and partly and more
interestingly because this can reveal new physics far from the standard
thermodynamic limit.

Let us first briefly review our present theoretical understanding of this
issue. In the canonical ensemble, heat capacities (proportional to the
fluctuations of the energy) are always positive. This ensemble is adequate
for (infinitely) large systems, with temperature driven by an external
thermal bath. In the microcanonical ensemble, the heat capacity at constant
volume is given by the derivative of the entropy $S$ with respect to the
energy $\frac{1}{c_vT^2}=-\frac{\partial^2 S}{\partial E^2}$, where
temperature is defined as $\frac{1}{T}=\frac{\partial S}{\partial E}$. This
is the correct ensemble for finite systems \cite{GROSS}. For these systems,
entropy is not necessarily an extensive function. Therefore, the entropy may
have a convex part as a function of the energy. If this occurs, then one
expects a backbending of the temperature as a function of the energy and a
negative heat capacity.  This effect should manifest most clearly
in ``small systems'', i.e. systems with a ``few'' number of particles
\footnote{ Systems with large number of particles interacting with long
range forces can also be considered \cite{GROSS,THI70}. This will not be
addressed here.}, due to  larger surface to volume ratios.

In practice, using data from experiment or from numerical simulations, two
equivalent methods have been used to calculate heat capacities. a) Directly,
using the definition $1/c_v=({\partial T}/{\partial E})_v$. One creates
samples at slightly different energies and calculates the corresponding
(microcanonical) temperatures. b) Using formula
\footnote{It should be noticed that in the
 derivation of this formula  \cite{85} one  considers 
 $\frac{\sigma_K^2}{\langle K\rangle^2}$ as a small parameter. Therefore,
 the formula is not valid  when $\sigma_K^2$ diverges.}

 \be c_v=\frac{C_1^2}{C_1-\frac{\sigma_K^2}{T^2}}
\label{e:sig},\ee
where $C_1=\frac{3}{2}N$ is the heat capacity of the perfect gas and 
 $\sigma_K^2=<K^2>-<K^2>$ the fluctuation of the kinetic energy.

A number of theoretical calculations predict the existence of negative heat
capacities in ``small systems''. Among the on-lattice calculations, we will
mention the microcanonical Metropolis Monte Carlo (MMC) results of
D.H.E. Gross \cite{GROSS} in a ten-colors Potts model, those of Chomaz and
Gulminelli {\cite{CHOMAZ,GULMINELLI} with a lattice-gas model with average
constrained volume and those of Pleimling and Hüller
\cite{PLEIMLING-HUELLER} with the $2-d$ and $3-d$ Ising model at constant
magnetization. In the domain of atomic clusters, Labastie and Whetten
\cite{LABASTIE}, using MMC techniques, predicted negative $c_v$ around the
melting temperature of small Lennard-Jones clusters. D.H.E. Gross
\cite{GROSS} who extensively investigated the microcanonical thermodynamics
of ``small systems'' found a clear signal of negative $c_v$ in realistic
calculations of the liquid-gas transition of metal clusters. Dorso and
collaborators also found a signal  for a Lennard-Jones fluid,
but only at very low densities (less than $1/10$ of the normal liquid
density) \cite{DORSO}.

The experimental finding of negative heat capacities in ``small systems''
has been announced, first in the fragmentation of atomic
nuclei \cite{DAGOSTINO} and latter in the solid-liquid transition of
sodium atoms clusters \cite{SCHMIDT} and in the liquid-gas transition of
hydrogen atoms clusters \cite{FARIZON}.

Method (a) has been used in references
\cite{GROSS,LABASTIE,CHOMAZ,DORSO,SCHMIDT,FARIZON} and method b) in references
\cite{CHOMAZ, DORSO}. The use of method b) is particularly difficult with
fast time evolving systems, because it demands a determination of
the fluctuations of the kinetic energy at times that are experimentaly
inaccesible.  The analysis of nuclear fragmentation data
\cite{DAGOSTINO,DAGOSTINO2} by D'Agostino et al. has been performed with a
method derived from b), adapted to the information given by the experimental
data.  Below we discuss in detail the validity of this method.

The goal is to infer for each event, the kinetic energy of the system when
fragmentation occurs. One proceeds as follows. This energy $K=E-V$ is first
written as the total energy E (taken as the sum of the binding $B_0<0$ and
the excitation $E^*$ energies of the system), minus the inter-particle
potential energy $V$.  Looking at the system as an ensemble of fragments,
the potential energy can be split into a sum $V=\sum_i V_i+\sum_{i<j}V_{ij}$
of intra-fragment $\sum_i V_i$ and inter-fragment potential energies
$\sum_{i<j}V_{ij}$. Adding and subtracting the sum of the binding energies
$B_i$ of the fragments and splitting the inter-fragment interaction into its
Coulomb and nuclear parts, one arrives at the (exact) expression

\be K=E-\sum_i (V_i-B_{i})-\sum_{i}B_{i}-\sum_{i<j}V_{ij}^{Coul}-
\sum_{i<j}V_{ij}^{nuc}
\label{e:K}\ee

In order to proceed with data it is assumed that the information taken at
late times, when fragments hit the detectors, suffices to reconstruct the
above energy partition at earlier times, when the system was at the required
thermodynamic conditions (of temperature, density, pressure...). In
practice, the quantities that are effectively measured are the total energy
$E$ and the charge of the fragments detected in the event. From the later,
with plausible hypothesis on the number and distribution of the undetected
neutrons, it is possible to estimate the value of $\sum_{i}B_{i}$ at early
times. This has been done carefully (see ref.\cite{DAGOSTINO,DAGOSTINO2})
and the authors have checked that the final results are rather insensitive
to details.  Next, it is assumed that when fragmentation takes place, the
volume of the system is sufficiently large (of the order of three to ten
times the normal volume) to neglect the nuclear interaction between the
fragments $\sum_{i,j}V_{ij}^{nuc}$.  The (inter-fragment) Coulomb energy
term is estimated placing the detected charge distribution of \it spherical
\rm and \it compact \rm fragments inside a sphere of a given volume.  The
amplitude of the fluctuations of this term is rather uncertain, because
neither the shape and compactness (see below) nor the positions of the
fragments inside the volume are under control. Finally, the term $\sum_i
(V_i-B_{i})$, which represents the change of intra-fragment potential energy
due to deformation and non-compactness, is neglected. With these
assumptions, the heat capacities are calculated from the fluctuations, at
fixed energy E, of the quantity \be
K'=E-\sum_{i}B_{i}-\sum_{i<j}V_{ij}^{Coul}
\label{e:K'}\ee
In short, the fluctuation of the kinetic energy $K$ is replaced by the
fluctuation of $K'$.  Data exhibit a significant rise of fluctuations of $K'$
in a domain of a few MeV/nucleon. Applying formula 
\be
c_v'=\frac{c_1^2}{c_1-\frac{\sigma_{K'}^2}{T^2}}
\label{r:K'}\ee
 (the
temperature is estimated assuming that fragments behave as classical
particles and nucleons inside fragments as a Fermi gas), the authors
\cite{DAGOSTINO,DAGOSTINO2} conclude that negative heat capacities have been
observed in nuclear fragmentation data.

It is rather straightforward to test in a simple model the effects of
substituting the fluctuations of $K$ by those of $K'$. We consider a
Lennard-Jones fluid with $N=64$ particles confined in a container.  In order
to better localize the origin of discrepancies, we take uncharged
particles. We will discuss below the possible influence of Coulomb
interactions.

Molecular dynamics equations of motion are solved with a time step $\delta
t=0.01(m\sigma^2/48\epsilon)^{1/2}$, ensuring a conservation of the total
energy better than $0.05$ percent. Technical details are as in reference
\cite{BIG-BANG}. Calculations (microcanonical) are performed at various
energies and volumes. Temperatures are calculated as $T=2K/3N$. We checked
that this is a good approximation of the true microcanonical temperature
\cite{85}. After a thermalization time of various $10^6$ time steps, the
relevant quantities are sampled every $10^3$ steps. Using autocorrelation
functions, we verified that these intervals are large enough. Averages are
made on samples of $10^4$ events. Fragment are identified using energetic
Hill's criterion \cite{HILL}. This method gives results very similar to
those of Dorso and Randrup \cite{DORSO-RANDRUP}. On-lattice, it is almost
identical to the Coniglio-Klein prescription, used by Chomaz and Gulminelli
\cite{CHOMAZ}.  From a mass table of Lennard-Jones clusters
\cite{LJ-CLUSTERS} we determine, event by event, the quantity
$K'=E-\sum_{i}B_{i}$ and following formula (4), we calculate $c'_{v}$.

\begin{figure}[h!] 
\begin{center}
\includegraphics[scale=0.5]{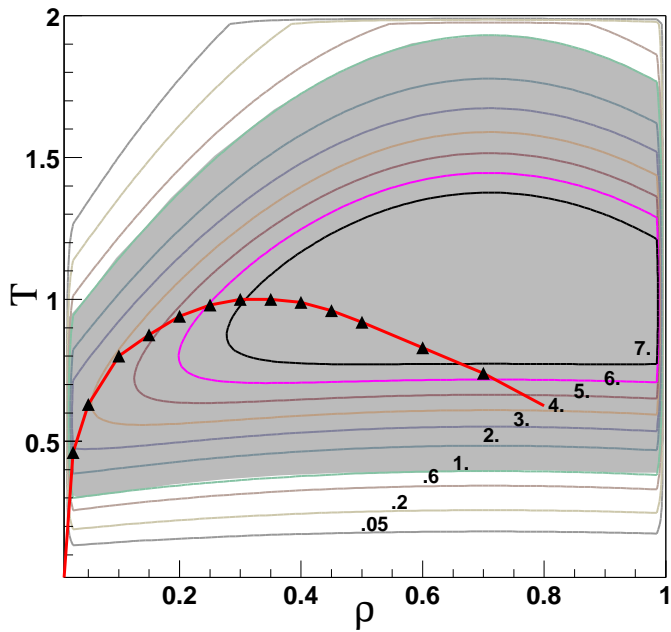}
\end{center}
\caption{\label{kline} \it Phase diagram of a Lennard-Jones fluid with
$N=64$ particles confined in a container. Iso-contour lines indicate values
of the quantity $\sigma_{K'}^2/T^2C_1$. When this quantity exceeds one (grey
area), $c_{v}'$ (see text) is negative. The position of the liquid-gas
coexistence line is sketched by filled triangles.  Temperature $T$ and
density $\rho$ are in units of the Lennard-Jones potential.}
\end{figure}

\begin{figure}[h!] 
\begin{center}
\includegraphics[scale=0.3]{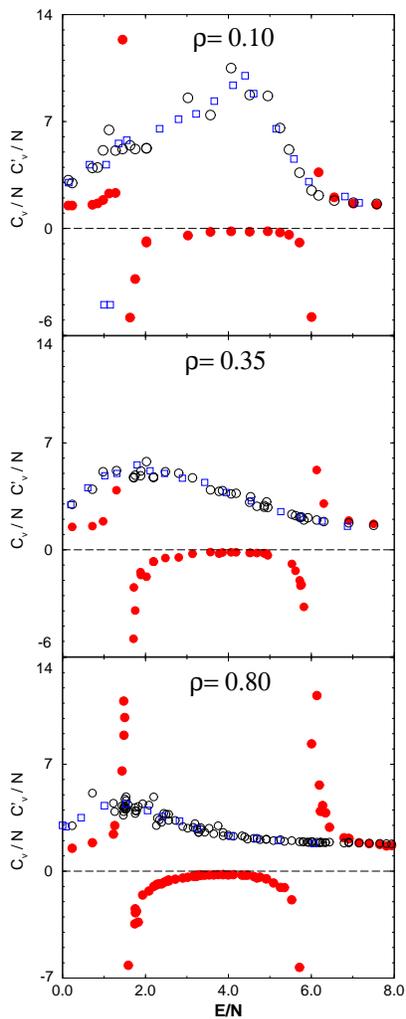}
\end{center}
\caption{\label{Cv} \it The heat capacity $c_v/N$ (open circles and open
squares, see text) and the quantity $c_v'/N$ (filled circles), as a function
of the excitation energy per particle $E^{*}/N$ (in units $\epsilon$ of the
Lennard-Jones potential) for isochore paths at $\rho=0.10, 0.35$ and
$0.80$.}
\end{figure}


We see that when $\sigma_{K'}^2/T^2C_1>1$, $c'_{v}$ becomes negative. Figure
1 reveals that this quantity exceeds one in a large zone of the
density-temperature phase diagram. The location of the zone of maximal
fluctuation of $K'$ is easily understood.  These fluctuations are nothing
but those of the fragment mass distribution $n(A)$ and one expects these
fluctuations to be maximum around the critical point. Indeed, if the binding
energy of the fragments is well represented by a mass formula
$B(A)=a_{v}A+a{s}A^{2/3}$, then $\sigma_{K'}^2=
a_{s}^2(<m_{2/3}^2>-<m_{2/3}>^2)$, where $m_{2/3}$ is the standard moment of
order $2/3$ of $n(A)$.

The above results show that, at least in the present model, $c'_{v}$ is not
a satisfactory representation of the heat capacity. Indeed, heat capacities
cannot be negative in the (monophasic) super-critical region
($T>T_{critical}$). Moreover, molecular dynamics simulations indicate that in
the present model $c_{v}$ is always positive (except in a small zone at very
low density as in ref.\cite{DORSO} and at low temperatures around the
liquid-solid transition as in ref. \cite{LABASTIE}). Examples of a
comparison of $c_{v}$ and $c'_{v}$ are given in Fig. 2 for isochore paths at
$\rho=0.1$, $\rho =0.35$ and $\rho =0.80$. The heat capacities $c_{v}$ have
been calculated from the fluctuations of $K$ (open circles) and from the
derivatives of the caloric curve (open squares). One remarks that both agree
very well. Note that for the density of $0.1$ and a narrow energy domain
around $1.0$, the later method predicts a negative heat capacity.

\begin{figure}[h!] 
\begin{center}
\includegraphics[angle=0,scale=0.5]{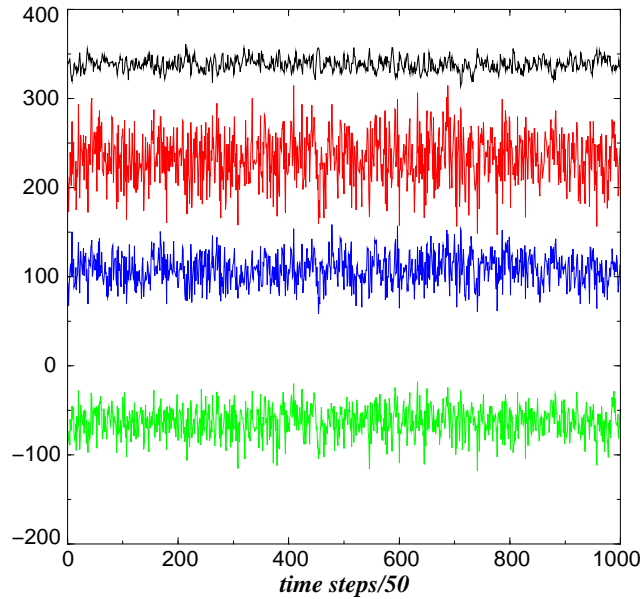}
\end{center}
\caption{\label{fluctuats1.ps} \it Time evolution of various quantities at a
density of $\rho=0.35$ and excitation energy per particle of
$E^{*}/N=4.12$. From top to bottom: $K$ (shifted by 250 units), $K'$ (shifted
by 100 units), $\sum_i (V_i-B_{i})$ and $\sum_{i<j} V_{i,j}$.  }
\end{figure}

\begin{figure}[h!] 
\begin{center}
\includegraphics[angle=0,scale=0.4]{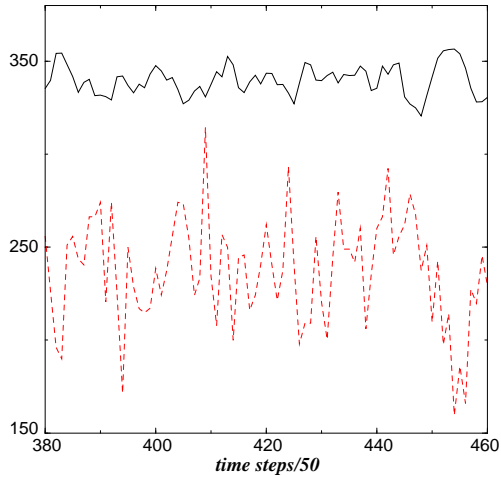}
\end{center}
\caption{\label{Fluctuats2.ps} \it Time evolution of $K$ (continuous line)
and $K'$ (dashed line), with an enlarged time scale. Same
conditions as for Fig.3}
\end{figure}

The origin of the discrepancy with the  conclusions of
Ref.\cite{DAGOSTINO,DAGOSTINO2} can be localized by looking at the various
components of K (equ.(2)) (see Fig.3 and 4). One immediately remarks that
the fluctuation of $K'$ largely exceeds the fluctuation of $K$. A closer
examination (see Fig.4) also reveals that these fluctuations are not in
phase. These discrepancies appear because the contributions of the terms
$\sum_i (V_i-B_{i})$ and $\sum_{i<j} V_{i,j}$ have been neglected. We have
observed that the contribution of the inter-fragment potential energy
decreases with density but surprisingly, the fluctuation of the
intra-fragment potential energy remains very important at all densities.
This means that in the regions of density and temperature we are looking at,
fragments are not \it compact spheres\rm. From the above considerations it
emerges that for the present simple model, where physics is well under
control, $c_{v}'$ is not a satisfactory approximation of the heat capacity.

The question that now naturally arises is to what extent this remains true
for real nuclei. One of the main differences is the absence of Coulomb
energy in our analysis. The motivation of our choice is that simpler is the
model, clearer is the identification of the sources of discrepancy. We do
not believe that including Coulomb interaction the situation would
drastically change. Ison and Dorso \cite{Ison} have shown that adding a
Coulomb term to a Lennard-Jones potential produces no qualitative changes in
the heat capacities of small systems. In any case, the problem of the
inter-fragment \cite{Bonasera} and intra-fragment energies will persist. By
the way, if Coulomb was so important in nuclei, this would mean that
the fluctuations are linked more to instabilities in a Coulomb gas than to a
liquid-gas phase transition and the whole interpretation should be revised.
Furthermore, one has to keep in mind that in the analysis of experimental
data \cite{DAGOSTINO,DAGOSTINO2} only part of the Coulomb fluctuations are
extracted from experiment, because at each event the spatial distribution of
fragments inside the fragmentation volume is unknown.

The present results illustrate once more the difficulty to study the
thermodynamics of  ``small systems'' from data that are (almost) limited
to fragment size distributions measured after their expansion at very late
times. We hope that the results presented here will stimulate the search for
new approaches of this problem.

We are indebted to Ph. Chomaz, F. Gulminelli and M. D'Agostino for fruitful
discussions and for pointing out an error in the plotting of Figures 3 and 4 of
a previous version of this paper. We would also like to thank D.H.E. Gross for
stimulating discussions on negative heat capacities in small systems.

\end{document}